\documentclass[aps,prd,twocolumn,superscriptaddress,nofootinbib,floatfix]{revtex4-2}

\usepackage{graphicx}
\usepackage{amsmath}
\usepackage{amssymb}
\usepackage{bm}
\usepackage{siunitx}
\usepackage{xcolor}
\usepackage{hyperref}
\usepackage{lineno}

\begin{document}

\title{GPU-accelerated spectrum reweighting for new-physics searches in solar neutrino--electron scattering}

\author{Guangbao Sun}
\affiliation{School of Physics and Technology, Wuhan University, Wuhan 430072, China}

\author{Xuefeng Ding}
\email{dingxf@ihep.ac.cn}
\affiliation{Institute of High Energy Physics, Chinese Academy of Sciences, Beijing 100049, China}

\author{Liang Sun}
\email{sunl@whu.edu.cn}
\affiliation{School of Physics and Technology, Wuhan University, Wuhan 430072, China}

\author{Xiang Zhou}
\email{xiangzhou@whu.edu.cn}
\affiliation{School of Physics and Technology, Wuhan University, Wuhan 430072, China}

\begin{abstract}
Precision measurements of neutrino--electron elastic scattering provide direct low-energy probes of weak interactions and of possible beyond-the-Standard-Model effects. Non-standard interactions (NSIs) and an anomalous neutrino magnetic moment modify the differential cross section through different kinematic structures, but both can change the normalization and spectral shape of the recoil-electron distribution. Likelihood-based tests of such effects become computationally demanding when each parameter point requires the recoil spectrum to be propagated through a detector-response model obtained from Monte Carlo (MC) simulation. We present a GPU-accelerated spectrum-reweighting framework that avoids regenerating detector MC samples for each new-physics parameter point. The method applies bin-to-bin weights at the recoil-spectrum level and propagates the reweighted spectrum through a fixed two-dimensional response model in recoil and reconstructed energy. This formulation retains the detector response inside the likelihood calculation while reducing each parameter update to operations on precomputed spectra and response kernels. The implementation uses NVIDIA \textsc{Thrust} transformation--reduction primitives and is compiled from a common source for CUDA and OpenMP back ends. For the benchmarks considered here, a single likelihood evaluation takes ${\sim}87$~ms on a commodity NVIDIA RTX~3080~Ti and ${\sim}52$~ms on a datacenter NVIDIA A30X; the latter is a $58\times$ speedup over a single CPU thread and ${\sim}2.5\times$ over a fully loaded 64-thread CPU. The consumer-GPU result, on hardware accessible to any user, shows that the framework supports interactive parameter scans on a single workstation. The principal acceleration, however, comes from avoiding regeneration of the detector Monte Carlo at each parameter point rather than from the GPU itself. The approach is applicable to neutrino--electron scattering analyses in which the new-physics dependence can be represented as a reweighting of an existing recoil spectrum, including the flavor-diagonal NSI and magnetic-moment examples considered in this work.
\end{abstract}

\maketitle

\section{Introduction}
\label{sec:intro}

Precision measurements of neutrino--electron elastic scattering provide a sensitive low-energy test of the Standard Model (SM) and of weakly coupled new physics. In the SM, the process receives charged- and neutral-current contributions whose relative weights depend on the incident neutrino flavor. Small deviations in the scattering cross section can therefore appear as changes in both the event rate and the shape of the recoil-electron spectrum. Spectral measurements of solar and reactor neutrinos consequently provide useful probes of beyond-the-Standard-Model (BSM) interactions.

Non-standard neutrino interactions (NSIs) provide a widely used effective description of such deviations. At low momentum transfer, NSIs with electrons can be represented by four-fermion operators with dimensionless couplings $\varepsilon_{\alpha\beta}^{eP}$ normalized to the Fermi constant $G_F$~\cite{Farzan:2017xzy,Dev:2019NSIstatus}. Depending on the underlying model, these operators may arise from heavy mediators in extended gauge or scalar sectors, or from light mediators with masses comparable to the momentum transfer in neutrino scattering~\cite{Dutta:2023light,Demirci:2024light}. In solar-neutrino experiments, NSIs can affect both propagation through the matter potential~\cite{Wolfenstein:1977ue,MSW:1986} and detection through the neutrino--electron cross section. This work focuses on the detection effect, namely cross-section modifications that distort the recoil spectrum for fixed solar fluxes and oscillation probabilities; recent solar-neutrino analyses have begun to place direct constraints on such detection-level NSIs~\cite{Borexino:2026nsi}.

The neutrino magnetic moment provides another example with a different physical origin but a similar computational structure. Its electromagnetic contribution to neutrino--electron scattering has a characteristic low-recoil enhancement proportional to $1/T$, where $T$ is the electron recoil kinetic energy~\cite{Giunti:2014ixa,Borexino:2017fbd,Beda:2012zz}. Although magnetic-moment interactions and NSIs correspond to different operators, they both modify the differential cross section before detector effects are applied. This common feature motivates a unified numerical strategy: compute the modified recoil spectrum and propagate it through the same detector-response model.

The framework is most suited to analyses in which new-physics hypotheses are tested through their distortions of reconstructed neutrino--electron recoil spectra. Completed solar-neutrino measurements such as SNO and Borexino illustrate the level of spectral modeling required to extract weak-interaction, oscillation, and beyond-the-Standard-Model information from recoil-electron data~\cite{SNO:2013cef,Borexino:2023CNO}, interpreted against precisely determined three-flavor oscillation parameters~\cite{NuFit6:2024,Xu:2023solar}. The same computational idea applies to present and future low-background measurements -- Super-Kamiokande, JUNO, and the Jinping Neutrino Experiment~\cite{SK:2024sol,JUNO:2025first,JNE:2017,Martinez:2022nsi}, and to reactor neutrino--electron searches including GEMMA-type magnetic-moment measurements~\cite{Beda:2012zz} -- wherever repeated spectral predictions are needed over NSI, magnetic-moment, or related BSM parameter spaces. The method is thus experiment-independent at the level of the response-folded recoil-spectrum calculation: it applies whenever the new-physics effect can be represented as a recoil-spectrum reweighting and the detector response is available for propagation to the reconstructed observable.

The main difficulty addressed in this work is computational. In a conventional likelihood analysis, the predicted recoil spectrum must be propagated through a detector-response model derived from full Monte Carlo (MC) simulation at each point in the new-physics parameter space. Repeating this operation over a multidimensional parameter space can dominate the cost of fits, scans, and Markov-chain Monte Carlo studies, especially when the detector response is represented by a two-dimensional mapping between true recoil energy and reconstructed observables.

We develop a GPU-accelerated spectrum-reweighting framework to reduce this cost. Instead of regenerating detector-level MC samples for every parameter point, the method starts from a baseline SM MC sample and incorporates new physics through bin-to-bin weights in the recoil-energy spectrum. In matrix notation, the predicted visible spectrum can be written as
\begin{equation}
  \bm{E}_{\rm obs}(\bm{\varepsilon})
    = \mathbf{R}\,\mathrm{diag}\bigl(\bm{r}(\bm{\varepsilon})\bigr)\,
      \bm{E}_{\rm true}^{\rm (SM)},
  \label{eq:matrix_reweight}
\end{equation}
where $\bm{\varepsilon}$ denotes the physics parameters being varied, $\bm{r}(\bm{\varepsilon})$ is the vector of bin-wise weights, $\bm{E}_{\rm true}^{\rm (SM)}$ is the baseline SM recoil spectrum, and $\mathbf{R}$ is the linear detector-response operator that maps a true recoil spectrum to the reconstructed-observable spectrum (its matrix form, with the index convention $i=$~true and $j=$~observable used throughout, is given in components in Eq.~\eqref{eq:visible_spectrum}, the operator contracting the true index $i$). The response matrix, which includes quenching, energy resolution, reconstruction effects, and other detector-specific effects, is computed once and then kept fixed. The new-physics dependence is thus isolated in the recoil-spectrum weights, while the detector response remains part of the prediction. Because the detector response for a recoil electron of energy $T$ is independent of the underlying interaction that produced it, $\mathbf{R}$ is unchanged under parameter variation; the only approximation introduced by the method is the bin-level evaluation of $\bm{r}(\bm{\varepsilon})$, whose accuracy we quantify in Sec.~\ref{sec:validation}.

Because the reweighting operation is parallel over spectrum and response-matrix bins, it is naturally suited to GPU acceleration. We implement the calculation with NVIDIA CUDA and the \textsc{Thrust} template library~\cite{Thrust}, using standard parallel patterns such as \texttt{transform}, \texttt{reduce}, and \texttt{transform\_reduce}. The same source can be compiled for CUDA and OpenMP back ends, allowing the GPU implementation to be checked against a CPU reference. This distinction is important: the cross-section formalism defines the physics prediction, whereas the \textsc{Thrust} implementation provides an efficient and reproducible way to evaluate that prediction repeatedly.

As reported in Sec.~\ref{sec:performance}, one likelihood call runs in tens of milliseconds on hardware ranging from a commodity NVIDIA RTX~3080~Ti to a datacenter NVIDIA A30X; these are not hardware-independent performance numbers but demonstrate that the spectrum-reweighting calculation maps efficiently onto GPUs across that range.

Several of the individual ingredients used here are established. GPU acceleration of neutrino analyses was demonstrated by Calland \textit{et al.}, who reweighted oscillation Monte Carlo event-by-event on a GPU for T2K~\cite{Calland:2013vaa,CallandCHEP2015}, and GPU evaluation of oscillation probabilities is provided by tools such as \textsc{CUDAProb3}~\cite{CUDAProb3}. Single-source frameworks that compile the same likelihood kernels for GPU and multicore CPU back ends, typically built on \textsc{Thrust}, are also well developed in particle physics, including \textsc{GooFit}~\cite{Andreassen:2013goofit,Schreiner:2018goofit} and \textsc{zfit}~\cite{Eschle:2019zfit}, while the \textsc{GNA} framework organizes a neutrino fit as a computational graph in which only parameter-dependent nodes are recomputed and the detector response is treated as a fixed transformation~\cite{Fatkina:2018gna}. At the statistical level, folding a parameter-dependent true spectrum through a fixed response matrix, with templates reweighted as parameters vary, is standard practice and is implemented in widely used fitting tools~\cite{Cranmer:2012histfactory,Heinrich:2021pyhf}. More broadly, GPU offloading is increasingly used for the most expensive steps of particle-physics simulation and analysis, for example detector simulation in \textsc{Celeritas}~\cite{Celeritas:2024} and matrix-element generation in \textsc{MadFlow}~\cite{MadFlow:2021}. The contribution of the present work is not any one of these elements in isolation, but their combination for solar neutrino--electron scattering: a bin-to-bin reweighting of the recoil spectrum folded through a fixed two-dimensional (recoil energy, reconstructed observable) detector-response matrix -- built from a full Geant4 simulation in the intended application, and validated here on a synthetic analytic response (Sec.~\ref{sec:validation}) so that no experimental data enter this methods study -- implemented as a single-source \textsc{Thrust} pipeline and applied to the flavor-diagonal NSI and neutrino-magnetic-moment cases. We make the factorization assumption and its accuracy explicit (Sec.~\ref{sec:validation}) and report timings together with their use-case framing (Sec.~\ref{sec:performance}).

The remainder of this paper is organized as follows. Section~\ref{sec:theory} reviews neutrino--electron scattering in the SM, with flavor-diagonal NSIs, and with a neutrino magnetic moment, and introduces the detector-response mapping. Section~\ref{sec:method} presents the spectrum-reweighting formalism using two-dimensional probability densities. Section~\ref{sec:software} describes the \textsc{Thrust}-based implementation and software architecture. Section~\ref{sec:performance} reports the benchmark setup and timing results. Section~\ref{sec:validation} verifies the reweighting against a direct calculation and demonstrates a representative parameter fit. Section~\ref{sec:conclusions} summarizes the scope and limitations of the framework.

\section{Neutrino--electron scattering and detector response}
\label{sec:theory}

\subsection{Standard Model $\nu$--$e$ elastic scattering}
\label{subsec:sm}

At solar-neutrino energies, elastic scattering on electrons is well described by an effective four-fermion interaction with vector and axial electron couplings $g_V^\alpha$ and $g_A^\alpha$ for an incident neutrino of flavor $\alpha=e,\mu,\tau$. The electron-neutrino channel receives both neutral- and charged-current contributions, whereas the $\mu$ and $\tau$ channels receive only the neutral current. In terms of the chiral combinations $g_{L,R}^\alpha = g_V^\alpha \pm g_A^\alpha$, the tree-level couplings are
\begin{align}
  g_L^e &= \frac{1}{2} + \sin^2\theta_W, &
  g_R^e &= \sin^2\theta_W, \\
  g_L^{\mu,\tau} &= -\frac{1}{2} + \sin^2\theta_W, &
  g_R^{\mu,\tau} &= \sin^2\theta_W,
\end{align}
where $\theta_W$ is the weak mixing angle.

Neglecting the neutrino mass and retaining the electron-mass terms relevant at solar energies, the differential cross section for $\nu_\alpha e^- \to \nu_\alpha e^-$ as a function of the electron recoil kinetic energy $T$ is
\begin{equation}
  \frac{d\sigma_{\alpha}^{\rm SM}}{dT}
    = \frac{2 G_F^2 m_e}{\pi}
  \biggl[
    g_L^{\alpha\,2}
    + g_R^{\alpha\,2}\left(1-\frac{T}{E_\nu}\right)^2
    - g_L^\alpha g_R^\alpha \frac{m_e T}{E_\nu^2}
  \biggr],
  \label{eq:sm_cross_section}
\end{equation}
where $E_\nu$ is the incident neutrino energy; see, for example, Refs.~\cite{Tomalak:2019ibg,Formaggio:2012cpf}.

Because of neutrino oscillations in vacuum and matter, solar-neutrino detectors observe a flavor mixture rather than a pure $\nu_e$ flux: for a solar component $k$ with production spectrum $\phi_k(E_\nu)$, the flux of flavor $\alpha$ at the detector is $\phi_k^\alpha(E_\nu) = P_{ek}^\alpha(E_\nu)\,\phi_k(E_\nu)$, with $P_{ek}^\alpha$ the energy-dependent $\nu_e\to\nu_\alpha$ transition probability. The SM prediction for the recoil-energy spectrum is then
\begin{equation}
  \frac{dR_{\rm SM}}{dT}
   = N_e \sum_{k,\alpha} \int dE_\nu\,
      \phi_k^\alpha(E_\nu)\,
      \frac{d\sigma_\alpha^{\rm SM}}{dT}(E_\nu,T),
  \label{eq:rate_sm}
\end{equation}
where $N_e$ is the number of target electrons.

\subsection{Flavor-diagonal nonstandard interactions}
\label{subsec:nsi_lagrangian}

At low energies, nonstandard interactions (NSIs) between neutrinos and electrons are described by an analogous effective four-fermion Lagrangian with dimensionless couplings $\varepsilon_{\alpha\beta}^{eP}$ normalized to $G_F$, where $P=L,R$ is the chirality of the electron current and $\alpha,\beta$ are flavor indices~\cite{Farzan:2017xzy,Dev:2019NSIstatus}. For the numerical demonstration in this work we restrict the analysis to flavor-conserving NSIs with real couplings ($\varepsilon_{\alpha\beta}^{eP}=0$ for $\alpha\neq\beta$), so that the NSI contribution is absorbed into shifts of the effective couplings,
\begin{equation}
  g_L^\alpha \to \tilde{g}_L^\alpha = g_L^\alpha + \varepsilon_{\alpha\alpha}^{eL},
  \qquad
  g_R^\alpha \to \tilde{g}_R^\alpha = g_R^\alpha + \varepsilon_{\alpha\alpha}^{eR},
  \label{eq:g_tilde}
\end{equation}
and the differential cross section retains the form of Eq.~\eqref{eq:sm_cross_section} with $g_{L,R}^\alpha\to\tilde g_{L,R}^\alpha$,
\begin{equation}
  \frac{d\sigma_{\alpha}^{\rm NSI}}{dT}
    = \left.\frac{d\sigma_{\alpha}^{\rm SM}}{dT}\right|_{g_{L,R}^\alpha\,\to\,\tilde g_{L,R}^\alpha}.
  \label{eq:xs_nsi}
\end{equation}
Complex or flavor-changing NSIs require the corresponding generalized cross section and flavor structure and are not included in the present examples. The corresponding recoil spectrum follows by replacing $d\sigma_\alpha^{\rm SM}/dT$ with $d\sigma_\alpha^{\rm NSI}/dT$ in Eq.~\eqref{eq:rate_sm}; for fixed solar fluxes and oscillation probabilities the NSI dependence enters entirely through the differential cross section at each phase-space point $(E_\nu,T)$, which is the key ingredient of the reweighting method introduced in Sec.~\ref{sec:method}.

\subsection{Neutrino magnetic moment}
\label{subsec:magneticmoment}

In addition to weak interactions, neutrinos may scatter off electrons through an anomalous magnetic moment $\mu_\nu$. In the minimally extended SM with massive Dirac neutrinos, a one-loop magnetic moment proportional to the neutrino mass arises, but it lies far below current experimental sensitivity; in many SM extensions it can be substantially enhanced, making it a useful probe of new physics~\cite{Fujikawa:1980yx,Shrock:1982sc,Giunti:2014ixa,Studenikin:2016ykv}.

Both the Dirac and Majorana cases are captured at the level of neutrino--electron scattering by an effective magnetic-moment parameter $\mu_\nu$, in units of the Bohr magneton $\mu_B=e/(2m_e)$, which we use in the following~\cite{Schechter:1981hw}.

The electromagnetic contribution to $\nu_\alpha e^- \to \nu_\beta e^-$ scattering proceeds through one-photon exchange and gives
\begin{equation}
  \frac{d\sigma^{\rm MM}}{dT}
    = \frac{\pi\alpha_{\rm em}^2}{m_e^2}
      \left(\frac{\mu_\nu}{\mu_B}\right)^{\!2}
      \left(\frac{1}{T} - \frac{1}{E_\nu}\right),
  \label{eq:xs_mm}
\end{equation}
where $\alpha_{\rm em}$ is the fine-structure constant and the remaining variables are defined as in Sec.~\ref{subsec:sm}; see, for example, Refs.~\cite{Vogel:1989iv,Wong:2005pa}. In this effective description, the magnetic-moment contribution is independent of the weak charges.

Two features of Eq.~\eqref{eq:xs_mm} are particularly relevant. First, the $1/T$ dependence enhances the cross section at low recoil energies and produces a spectral distortion that differs from the SM weak contribution. Low-threshold solar-neutrino detectors are therefore especially sensitive to $\mu_\nu$. Second, the electromagnetic amplitude flips the neutrino helicity, so the magnetic-moment contribution does not interfere with the SM weak amplitude, and the total differential cross section is an incoherent sum,
\begin{equation}
  \frac{d\sigma_\alpha^{\rm tot}}{dT}
    = \frac{d\sigma_\alpha^{\rm SM}}{dT}
      + \frac{d\sigma^{\rm MM}}{dT},
  \label{eq:xs_total}
\end{equation}
with $d\sigma_\alpha^{\rm SM}/dT$ given by Eq.~\eqref{eq:sm_cross_section}. If a magnetic moment is considered together with NSIs, the weak term in Eq.~\eqref{eq:xs_total} should be replaced by the corresponding NSI-modified cross section of Eq.~\eqref{eq:xs_nsi}.

Within the reweighting framework of Sec.~\ref{sec:method}, the magnetic moment enters through the same bin-level reweighting as the NSI case: because $d\sigma^{\rm MM}/dT$ depends only on the kinematic variables $(E_\nu,T)$ that define the recoil spectrum, the incoherent sum of Eq.~\eqref{eq:xs_total} multiplies each recoil bin by $1+(d\sigma^{\rm MM}/dT)/(d\sigma_{\alpha_i}^{\rm SM}/dT)$, leaving the computational structure unchanged. In practice the $1/T$ behavior in Eq.~\eqref{eq:xs_mm} is regulated by the finite analysis threshold and by the finite representative energy $T_i>0$ assigned to each recoil bin.

The strongest laboratory constraints on $\mu_\nu$ come from neutrino--electron scattering experiments. Borexino has reported an upper limit of $\mu_\nu < 2.8\times 10^{-11}\,\mu_B$ at 90\% C.L.\ using $^7$Be solar neutrinos~\cite{Borexino:2017fbd}, while GEMMA obtained a comparable bound of $\mu_\nu < 2.9\times 10^{-11}\,\mu_B$~\cite{Beda:2012zz}. Future solar-neutrino experiments with lower thresholds and improved energy resolution may further strengthen these constraints~\cite{SK:2024low}. This motivates including $\mu_\nu$ as a second physics example in the reweighting framework.

\subsection{Detector response and visible energy}
\label{subsec:detector_response}

Solar-neutrino detectors do not directly measure the recoil kinetic energy $T$ of the scattered electron. Instead, the observable is a reconstructed, or \emph{visible}, energy $E_{\rm vis}$ that differs from $T$ because of detector effects. Modeling the mapping $T \to E_{\rm vis}$ is therefore essential for connecting the theoretical cross sections in Secs.~\ref{subsec:sm}--\ref{subsec:magneticmoment} to the measured spectrum.

In liquid-scintillator detectors such as Borexino and JUNO, this mapping is governed by scintillation quenching (a non-proportional light yield~\cite{Birks:1951}), Cherenkov light, photoelectron statistics, position-dependent light collection, electronics response, and reconstruction. These effects cannot be captured by a single analytic expression, so a full detector Monte Carlo (MC) simulation is the standard modeling tool.

\paragraph{Geant4-based detector simulation and 2D mapping.}
In our framework, the detector response is derived from a Geant4-based simulation~\cite{GEANT4:2002zbu}. Each recoil electron with true kinetic energy $T$ is propagated through the detector geometry, including optical photon transport, PMT response, and event reconstruction. For each simulated event, we record the pair $(T,E_{\rm vis})$, where $E_{\rm vis}$ is the reconstructed visible energy obtained from the same reconstruction chain used for data. The ensemble of events defines a two-dimensional mapping, or equivalently a detector response matrix $\mathbf{R}$, that encodes the conditional probability
\begin{equation}
  P(E_{\rm vis}\,|\,T).
\end{equation}
For a given physics model, the predicted visible-energy spectrum is obtained by applying $\mathbf{R}$ to the true recoil-energy spectrum,
\begin{equation}
  \left.\frac{dR}{dE_{\rm vis}}\right|_j
   = \sum_i R_{ij}\,\left.\frac{dR}{dT}\right|_i,
  \label{eq:visible_spectrum}
\end{equation}
where $R_{ij}\equiv P(E_{{\rm vis},j}\,|\,T_i)$ is the \emph{conditional} detector response, with $i$ the true-recoil bin and $j$ the reconstructed-observable bin, so that $\sum_j R_{ij}$ equals the (energy-dependent) detection efficiency. As described in Sec.~\ref{subsec:2d_reweight}, the matrix is stored operationally in an unnormalized (joint) form built directly from the simulated $(T,E_{\rm vis})$ pairs, from which this conditional response is recovered.

\paragraph{Implication for reweighting.}
The detector response depends on the recoil-electron kinematics but not on the microscopic interaction that produced the electron: an electron of kinetic energy $T$ generates the same scintillation and Cherenkov signals whether it originates from the SM weak interaction, an NSI contribution, or a magnetic moment. A single baseline MC sample generated under the SM hypothesis therefore fixes the mapping between $T$ and $E_{\rm vis}$; when the physics parameters vary, only the bin-to-bin recoil weights are updated while this response is held fixed, retaining the Geant4-based detector model with the new-physics dependence in a factorized form suited to efficient GPU evaluation.

\section{Reweighting method}
\label{sec:method}

\subsection{Recoil-energy spectra and the reweighting function}
\label{subsec:weights}

The reweighting procedure starts from the recoil-energy spectrum of each neutrino component $k$. For a continuous-spectrum source, such as $pp$, $^8$B, and CNO neutrinos, the predicted differential rate in a recoil-energy bin $T_l$ can be written as
\begin{equation}
  f_T^{(k)}(T_l;\bm{\varepsilon})
    = \sum_j \phi_k(E_{\nu,j})\,
      \frac{d\sigma}{dT}(E_{\nu,j},T_l;\bm{\varepsilon})\,\Delta E_\nu,
  \label{eq:fT}
\end{equation}
where the sum runs over a discretized neutrino-energy grid $\{E_{\nu,j}\}$, $\phi_k(E_{\nu,j})$ denotes the flux spectrum after oscillation effects are included, and $d\sigma/dT$ is the differential cross section evaluated at the parameter point $\bm{\varepsilon}$. For solar neutrinos the detector sees a flavor mixture, so the product $\phi_k\,d\sigma/dT$ is understood as the sum over contributing flavors $\alpha$ of $\phi_k^\alpha\,d\sigma_\alpha/dT$ [cf.\ Eq.~\eqref{eq:rate_sm}], with the flavor-dependent (NSI-shifted) couplings of Eq.~\eqref{eq:g_tilde}. Here and below, $\bm{\varepsilon}$ denotes the set of new-physics parameters varied in the scan; for a magnetic-moment-only scan it may be replaced by $\mu_\nu$. For monoenergetic components, such as $^7$Be and $pep$, the sum reduces to a single contribution at fixed $E_\nu$. The SM reference spectrum $f_T^{{\rm SM},(k)}(T_l)$ is obtained by setting all new-physics parameters to zero.

The central idea of the method is to avoid regenerating the detector Monte Carlo (MC) sample whenever the physics parameters are updated. Instead, we generate one baseline sample under the SM hypothesis and encode the new-physics dependence through a multiplicative reweighting function. To this end, we define the normalized SM recoil-energy distribution for component $k$ as
\begin{equation}
  \hat{f}_T^{{\rm SM},(k)}(T_l)
    = \frac{f_T^{{\rm SM},(k)}(T_l)}
           {\displaystyle\sum_{l'} f_T^{{\rm SM},(k)}(T_{l'})},
  \label{eq:fT_norm}
\end{equation}
and introduce
\begin{equation}
  r^{(k)}(T_l;\bm{\varepsilon})
    = \frac{f_T^{(k)}(T_l;\bm{\varepsilon})}
           {\hat{f}_T^{{\rm SM},(k)}(T_l)}.
  \label{eq:reweight_factor}
\end{equation}
Here $r^{(k)}$ is the reweighting function constructed relative to the normalized SM reference spectrum. In this convention, the parameter-dependent recoil spectrum enters through $r^{(k)}$, while the final reconstructed distribution is normalized explicitly in the next step. All dependence on oscillation probabilities, modified cross sections, and the neutrino magnetic moment enters through the numerator of Eq.~\eqref{eq:reweight_factor}; the denominator is computed once during initialization and then kept fixed throughout the fit.

\subsection{Two-dimensional MC reweighting}
\label{subsec:2d_reweight}

The connection between the theoretical recoil spectrum and the experimentally measured observable is encoded in a two-dimensional MC histogram $N_{ij}^{(k)}$, where $i$ labels the recoil-energy bin $T_i$ and $j$ labels the reconstructed-observable bin. In the present analysis, this observable is the number of detected photoelectrons, $N_p$. The histogram $N_{ij}^{(k)}$ is obtained from a Geant4-based detector simulation for component $k$ under the SM hypothesis and therefore includes the detector response relevant to this channel: scintillation quenching, Cherenkov light, photoelectron statistics, and reconstruction effects.

From $N_{ij}^{(k)}$, we construct the response matrix
\begin{equation}
  R_{ij}^{(k)}
    = \frac{N_{ij}^{(k)} / M_j^{(k)}}
           {\displaystyle\sum_{i',j'} N_{i'j'}^{(k)} / M_{j'}^{(k)}}
      \times s^{(k)},
  \label{eq:Rij}
\end{equation}
where $M_j^{(k)}$ is a live-time correction mask that accounts for bin-dependent detection efficiency (the superscript $(k)$ is dropped in Eq.~\eqref{eq:mu_total}, where a single mask common to all components is applied), and $s^{(k)}$ is a scale factor chosen so that the integral of $R_{ij}^{(k)}$ matches that of the corresponding one-dimensional MC probability density function (PDF) in the analysis region of interest. Both $N_{ij}^{(k)}$ and $R_{ij}^{(k)}$ are computed once during initialization and stored as device-resident arrays. Note that $R_{ij}^{(k)}$ defined here is the SM \emph{joint} distribution of $(T_i,j)$, normalized globally rather than per true bin; it is therefore not identical to the conditional response $P(E_{\rm vis}|T)$ of Eq.~\eqref{eq:visible_spectrum}. The conditional transfer kernel is recovered by dividing by the normalized SM recoil reference $\hat{f}_T^{{\rm SM},(k)}(T_i)$, as done explicitly in Eq.~\eqref{eq:fNp_reweight} below.

At an arbitrary parameter point $\bm{\varepsilon}$, the predicted $N_p$ distribution for component $k$ is obtained by contracting the response matrix with the reweighted recoil spectrum,
\begin{equation}
  f_{N_p}^{(k)}(j;\bm{\varepsilon})
    = \frac{1}{\displaystyle\sum_{l} f_T^{(k)}(T_l;\bm{\varepsilon})}
      \sum_i \frac{R_{ij}^{(k)}}
                  {\hat{f}_T^{{\rm SM},(k)}(T_i)}\,
             f_T^{(k)}(T_i;\bm{\varepsilon}).
  \label{eq:fNp_reweight}
\end{equation}
The prefactor normalizes the resulting distribution. Inside the sum, the combination
$R_{ij}^{(k)}/\hat{f}_T^{{\rm SM},(k)}(T_i)$ acts as a transfer kernel from recoil energy to the reconstructed observable, while $f_T^{(k)}(T_i;\bm{\varepsilon})$ carries the new-physics dependence. When $\bm{\varepsilon}=0$, the SM prediction is recovered by construction.

For background components, such as $^{210}$Bi, $^{85}$Kr, and $^{11}$C, the spectra are independent of the new-physics parameters considered here. Their one-dimensional MC PDFs can therefore be used directly without reweighting.

\begin{figure*}[!htb]
  \centering
  \includegraphics[width=0.9\textwidth]{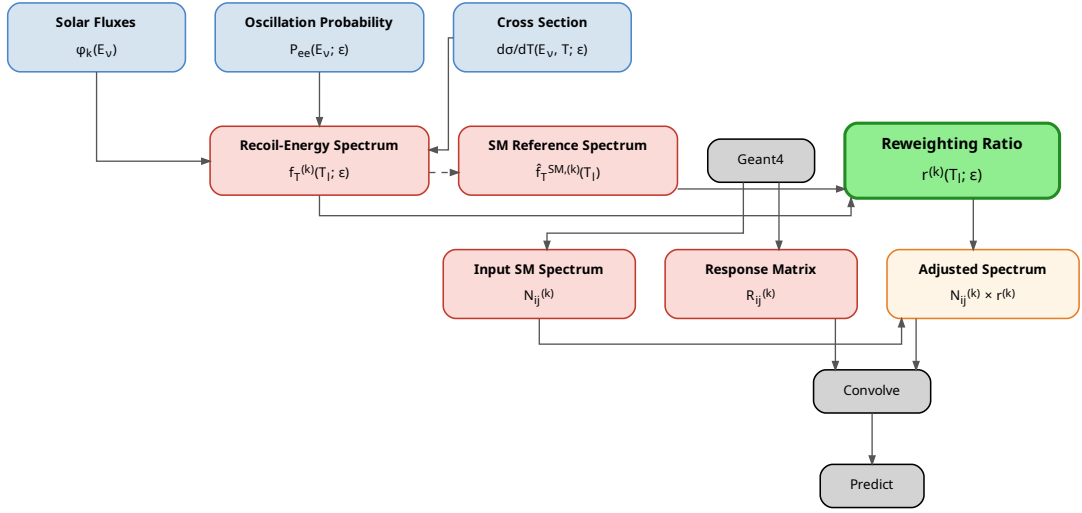}
  \caption{Schematic overview of the reweighting workflow. The theoretical calculation combines the oscillated solar fluxes $\phi_k(E_\nu)$ and the differential cross section $d\sigma/dT$ to construct the recoil-energy spectrum $f_T^{(k)}$ and the corresponding normalized SM reference $\hat{f}_T^{\rm SM}$. These quantities define the reweighting function $r^{(k)} = f_T^{(k)}/\hat{f}_T^{\rm SM}$. A Geant4 simulation under the SM hypothesis provides the two-dimensional histogram $N_{ij}^{(k)}$, from which the response matrix $R_{ij}^{(k)}$ is derived. Applying the reweighting function and the response matrix yields the predicted reconstructed spectrum $f_{N_p}^{(k)}(j;\bm{\varepsilon})$.}
  \label{fig:workflow}
\end{figure*}

\subsection{Predicted spectrum and likelihood}
\label{subsec:likelihood}

The total predicted spectrum is obtained by summing over all signal and background components,
\begin{equation}
  \mu_j(\bm{\varepsilon},\bm{\theta})
    = \text{MT} \times M_j
      \sum_k \text{rate}_k(\bm{\theta})\,
             f_{N_p}^{(k)}(j;\bm{\varepsilon}),
  \label{eq:mu_total}
\end{equation}
where $\text{MT}$ is the mass--time exposure, $M_j$ is the live-time mask for bin $j$, and $\text{rate}_k$ denotes the expected event rate of component $k$. For signal components, this rate depends on the solar-neutrino flux and the integrated cross section; for background components, it depends on the corresponding activity or external constraint. The vector $\bm{\theta}$ denotes nuisance parameters, including flux normalizations and background rates.

For binned data with observed counts $n_j$, we use the Poisson log-likelihood
\begin{equation}
  \begin{aligned}
    -2\ln\mathcal{L}(\bm{\varepsilon},\bm{\theta})
     &= 2\sum_j \Bigl[
         \mu_j - n_j \ln \mu_j + \ln\Gamma(n_j+1)
       \Bigr] \\
    &\quad + \sum_m \frac{(\theta_m - \hat\theta_m)^2}{\sigma_m^2},
  \end{aligned}
  \label{eq:logL}
\end{equation}
where the second term introduces Gaussian constraints on nuisance parameters with central values $\hat\theta_m$ and uncertainties $\sigma_m$. The fit is performed by minimizing Eq.~\eqref{eq:logL} with respect to $\bm{\varepsilon}$ and $\bm{\theta}$, either with \textsc{Minuit}~\cite{James:1975dr} in C++ or through \textsc{iminuit} in Python, in both cases using the \textsc{migrad} algorithm. Because the GPU and OpenMP back ends are compiled from a common source, a given fit returns the same minimum on either device up to floating-point rounding.

The computational cost of each likelihood evaluation is dominated by the reweighting operation in Eq.~\eqref{eq:fNp_reweight}, which must be repeated for each neutrino component and each parameter point. Because the sum over the recoil-energy index $i$ is independent for each reconstructed-energy bin $j$, the calculation is naturally parallel and maps efficiently onto the \textsc{Thrust} transform--reduce pipeline described in Sec.~\ref{sec:software}.

\section{Software implementation and architecture}
\label{sec:software}

\subsection{Design overview and \textsc{Thrust} abstraction}
\label{subsec:design_goals}

The reweighting formalism described in Sec.~\ref{sec:method} is implemented with NVIDIA's \textsc{Thrust} template library~\cite{Thrust}. In our framework, its parallel primitives (\texttt{transform}, \texttt{reduce}, and \texttt{transform\_reduce}) express the analysis chain from the evaluation of oscillation probabilities and differential cross sections to the construction of the predicted spectrum and likelihood.

All physics kernels, including the evaluation of $P_{ee}$, $d\sigma/dT$, recoil-energy spectra, detector-response propagation, and likelihood terms, are implemented as \texttt{\_\_host\_\_ \_\_device\_\_} functions and passed to \textsc{Thrust} algorithms through device lambdas. Because \textsc{Thrust} supports multiple execution back ends, the same source code can be compiled for CUDA on GPUs and OpenMP on CPUs by changing a single CMake option. This design avoids low-level device-specific kernel code while preserving portability across hardware platforms.

To improve memory locality and parallel efficiency, we adopt a structure-of-arrays layout. For each neutrino component $k$, separate arrays are maintained for the neutrino-energy grid $\{E_{\nu,j}\}$, oscillated flux spectrum $\{\phi_j\}$, survival probability $\{P_{ee,j}\}$, differential cross section $\{d\sigma/dT\}_{jl}$, recoil-energy spectrum $\{f_T\}_l$, normalized SM reference spectrum $\{\hat{f}_T^{\rm SM}\}_l$, and detector-response matrix $\{R_{ij}\}$. These arrays are initialized once and remain resident on the target device throughout the fit, minimizing host--device data transfer during repeated likelihood evaluations.

\subsection{\textsc{Thrust}-based transform--reduce pipeline}
\label{subsec:pipeline}

A single likelihood evaluation is expressed as a sequence of \textsc{Thrust} transformations and reductions that mirrors the formalism of Sec.~\ref{sec:method}. The pipeline consists of three stages: the physics update, the MC reweighting step, and the final spectrum and likelihood evaluation.

\paragraph{Physics update.}
For a given parameter vector $\bm{\varepsilon}$, the first stage computes the updated recoil-energy spectra for all neutrino components. A sequence of \texttt{transform} operations evaluates the oscillation probabilities and differential cross sections on the discretized kinematic grid, and then combines them with the oscillated fluxes to obtain $f_T^{(k)}(T_l;\bm{\varepsilon})$ as defined in Eq.~\eqref{eq:fT}. For composite components, such as CNO or the two $^7$Be lines, the corresponding subcomponents are combined through an additional weighted transformation.

\paragraph{MC reweighting.}
The second stage maps the updated recoil-energy spectrum to the reconstructed observable using the response matrix. For each neutrino component, a \texttt{transform} over the reconstructed-energy bin index evaluates Eq.~\eqref{eq:fNp_reweight}. Within each transformation, the sum over recoil-energy bins applies the precomputed response kernel to the updated spectrum and returns the corresponding reconstructed distribution. Background components are handled separately because their PDFs do not depend on the new-physics parameters and can be read directly from precomputed one-dimensional templates.

\paragraph{Spectrum accumulation and likelihood evaluation.}
In the final stage, all signal and background contributions are accumulated to form the predicted total spectrum $\mu_j$. The Poisson log-likelihood of Eq.~\eqref{eq:logL} is then evaluated through a fused \texttt{transform\_reduce} operation, in which each reconstructed-energy bin contributes its local likelihood term and the full sum is reduced to a single scalar. This fused implementation reduces intermediate memory traffic and limits host--device communication to the transfer of the final likelihood value; Gaussian pull terms for the nuisance parameters are added on the CPU side. The pipeline thus makes the computational bottleneck explicit -- the repeated application of the response kernel during parameter scans -- with each reconstructed-energy bin processed independently.

\subsection{Program architecture and interface}
\label{subsec:architecture}

\begin{figure*}[!htb]
  \centering
  \includegraphics[width=0.9\textwidth]{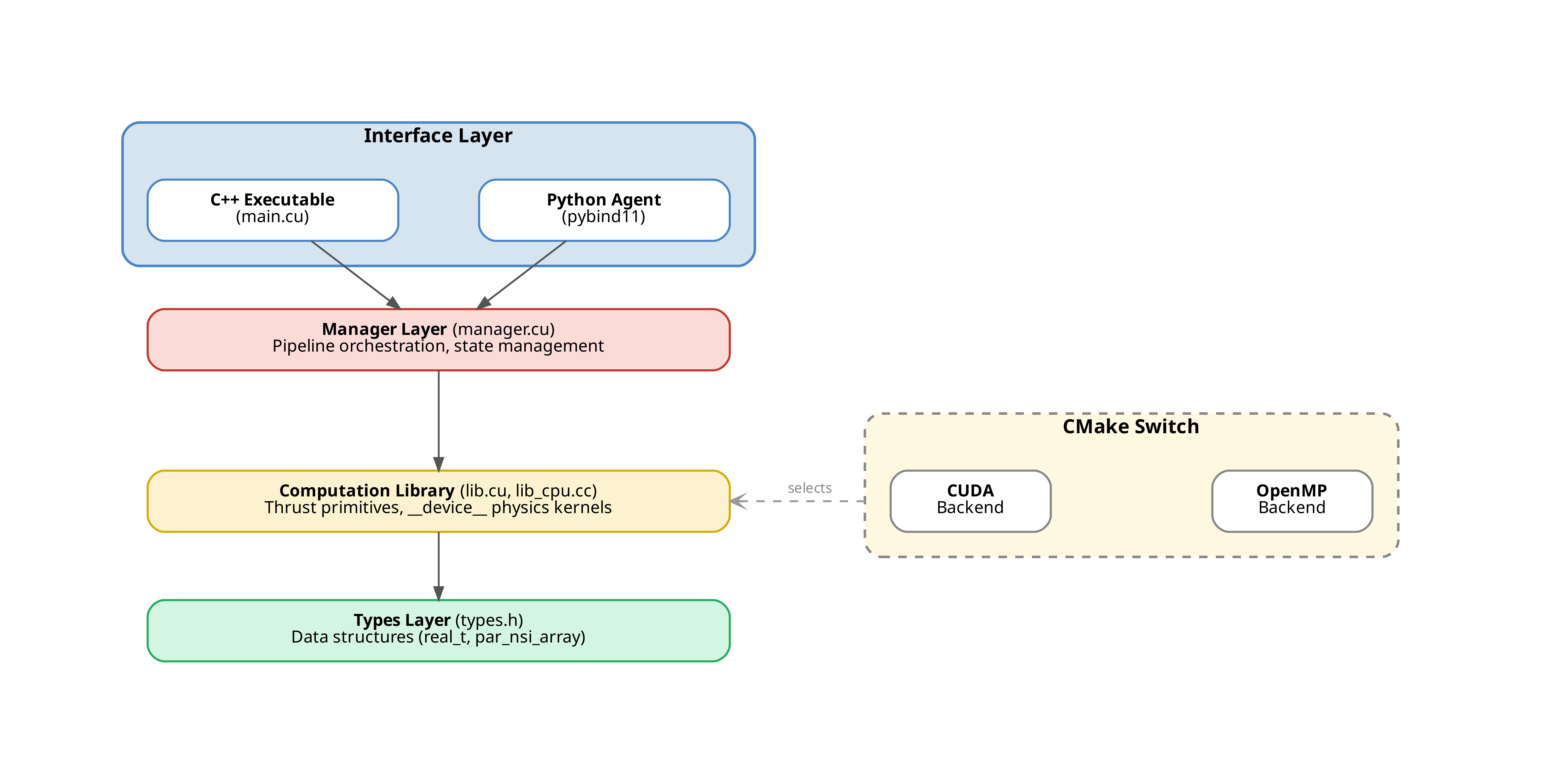}
  \caption{Software architecture of the reweighting framework. The code is organized into four layers: types, computation, manager, and interface. The same source can be compiled against either the CUDA or OpenMP back end through a CMake configuration switch.}
  \label{fig:architecture}
\end{figure*}

The software is organized into four layers, each built as a separate library with explicit CMake dependencies (Fig.~\ref{fig:architecture}), separating physics definitions, numerical kernels, workflow orchestration, and user interface. The \emph{types} layer fixes the numerical precision through the alias \texttt{real\_t} (\texttt{float} or \texttt{double} at compile time) and the data structures holding the NSI couplings, magnetic-moment parameter, and other analysis parameters. The \emph{computation} layer contains the core numerical routines -- oscillation probabilities, cross sections, detector-response propagation, and likelihood terms -- together with the \textsc{Thrust}-dispatched host routines and the loading of ROOT histograms into device-resident arrays. The \emph{manager} layer owns the analysis state: it loads inputs, constructs the SM reference spectra, initializes the response matrices, updates parameter-dependent quantities, and returns the likelihood for a requested parameter point, bridging the formalism and the numerical implementation. The \emph{interface} layer exposes the framework to end users, providing the C++ minimization callbacks and a \textsc{pybind11} Python wrapper so that the same pipeline is accessible from standalone C++ workflows and Python-based fitting tools.

\section{Performance evaluation}
\label{sec:performance}

\subsection{Benchmark setup}
\label{subsec:benchmark_setup}

We benchmark two configurations. \emph{Workstation~A} (datacenter) pairs a single NVIDIA A30-class GPU (reported by the driver as \texttt{A30X}; Ampere, compute capability 8.0, 24~GB HBM2; one of two installed, idle during the measurement) with a dual-socket Intel Xeon Platinum 8358P CPU (Ice Lake, 64 physical cores / 128 threads at 2.60~GHz), under Ubuntu~24.04.4 (kernel 6.8.0-90), CUDA~12.8 (driver 580.126.09), GCC~13.3.0, CMake~3.28.3, and ROOT~6.26.06. \emph{Workstation~B} (commodity, rentable on a public cloud) pairs a consumer NVIDIA GeForce RTX~3080~Ti GPU (Ampere, compute capability 8.6) with an Intel Xeon Silver 4214R CPU, under Ubuntu~22.04, CUDA~12.8, GCC~11.4, and ROOT~6.30.08. Both build the CUDA and OpenMP back ends from identical source (C++14) through the corresponding CMake configuration; the CPU thread-scaling reported below is measured on Workstation~A.

The reported timings correspond to a single likelihood evaluation, defined as one complete execution of the three-stage pipeline of Sec.~\ref{sec:software}. All results are reported in double precision. Each configuration was timed over ten consecutive likelihood evaluations with the first excluded as warm-up; Table~\ref{tab:timing} quotes the mean and standard deviation of the remaining nine.

The benchmark exercises the full multi-component analysis configuration of Sec.~\ref{sec:method}: the solar-neutrino signal and radioactive-background components, each folded through the two-dimensional recoil--$N_p$ detector response of Sec.~\ref{subsec:2d_reweight} on the framework's default $2000$-bin recoil grid. Because each likelihood call performs the same fixed sequence of array operations irrespective of the numerical contents of the spectra and the response matrix, the per-call latency depends only on these array dimensions and not on the underlying data. The timing is therefore reproducible from representative inputs at the production dimensions and requires no experimental detector sample, consistent with the data-independent scope of this methods paper.

\subsection{Benchmark results}
\label{subsec:results}

Table~\ref{tab:timing} summarizes the wall-clock time for one likelihood evaluation on the tested hardware. On the CPU, the OpenMP implementation scales with thread count and reaches $273\pm4$~ms at 16 threads, an $11.2\times$ speedup over a single thread. Beyond this the gain saturates ($132\pm2$~ms at 64 threads, $23.2\times$), consistent with finite memory bandwidth, synchronization overhead, and parallel-scheduling costs on the tested platform.

On the GPU, the A30X completes the same likelihood evaluation in $52.4\pm0.5$~ms, a $58\times$ speedup over a single CPU thread and a factor of $2.5$ relative to the best (64-thread) CPU configuration. On the commodity RTX~3080~Ti (Workstation~B) the same evaluation takes $86.8\pm1.2$~ms -- only ${\sim}1.7\times$ the datacenter time -- so the framework runs at interactive speed on hardware accessible to any user. These numbers should be interpreted as benchmark results for the stated hardware and software configurations, not as universal speedups for all detector models or parameter scans.

The GPU-versus-CPU comparison is, however, not the main source of acceleration. The algorithmic gain of the method is that it avoids regenerating the detector-level Monte Carlo sample at each parameter point: the alternative of propagating a freshly generated recoil spectrum through the full Geant4 simulation requires a dedicated detector-simulation campaign per parameter point -- minutes to hours of CPU or GPU time depending on the target statistics -- so reducing each parameter update to the ${\sim}52$~ms reweighting evaluated here is a speedup of many orders of magnitude relative to that brute-force baseline, independent of the GPU/CPU choice. The GPU acceleration is a further, smaller factor on top of this algorithmic gain.

\begin{table}[!htb]
  \centering
  \caption{Wall-clock time for a single likelihood evaluation in double precision (mean $\pm$ standard deviation over nine evaluations, first excluded as warm-up). Speedups are relative to the single-thread CPU result on Workstation~A; the commodity RTX~3080~Ti is measured on Workstation~B (absolute time only).}
  \label{tab:timing}
  \begin{tabular}{lcc}
    \hline
    Device / Configuration & Time (ms) & Speedup \\
    \hline
    \multicolumn{3}{l}{\textit{NVIDIA RTX 3080 Ti GPU (commodity, Workstation B)}} \\
    \quad CUDA               & $86.8\pm1.2$ & --            \\
    \hline
    \multicolumn{3}{l}{\textit{NVIDIA A30X GPU (Workstation A)}} \\
    \quad CUDA               & $52.4\pm0.5$ & $58.4\times$  \\
    \hline
    \multicolumn{3}{l}{\textit{Xeon Platinum 8358P (Workstation A, 64 cores)}} \\
    \quad OpenMP, 1 thread   & $3058\pm5$   & $1\times$     \\
    \quad OpenMP, 4 threads  & $943\pm116$  & $3.2\times$   \\
    \quad OpenMP, 16 threads & $273\pm4$    & $11.2\times$  \\
    \quad OpenMP, 32 threads & $146\pm1$    & $20.9\times$  \\
    \quad OpenMP, 64 threads & $132\pm2$    & $23.2\times$  \\
    \hline
  \end{tabular}
\end{table}

\begin{figure}[!htb]
  \centering
  \includegraphics[width=\linewidth]{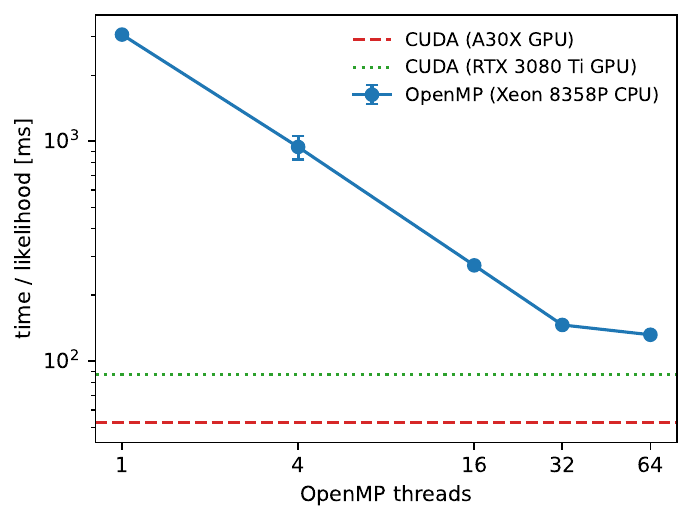}
  \caption{Wall-clock time per likelihood evaluation as a function of the number of OpenMP threads on the Xeon Platinum 8358P CPU (Workstation~A; error bars: standard deviation over nine evaluations). The dashed and dotted horizontal lines indicate the CUDA results on the A30X (Workstation~A) and the commodity RTX~3080~Ti (Workstation~B) GPUs, respectively.}
  \label{fig:timing_scaling}
\end{figure}

For iterative fits or scans, the total wall-clock time is the product of the per-call latency and the number of likelihood evaluations required by the optimizer or sampler, plus initialization overhead. Any claim about total fit time would therefore need to specify the minimizer, parameter space, stopping criterion, and number of likelihood calls; here we report only the reproducible per-call latency and a representative fit in Sec.~\ref{sec:validation}. The present benchmark shows that the dominant per-call operation can be evaluated rapidly on a single device without replacing the detector response by a simplified surrogate model.

\section{Validation and demonstration}
\label{sec:validation}

This section establishes that the reweighting reproduces the spectrum it is meant to replace (Sec.~\ref{subsec:closure}) and that the reweighting method runs through a complete parameter scan end to end (Sec.~\ref{subsec:demo}). The tests here validate the method and its numerical accuracy; the per-call speed of its GPU implementation is reported separately in Sec.~\ref{sec:performance}. To keep the validation independent of any particular detector simulation or data set, both tests are carried out on a fully analytic problem for which the exact answer is known in closed form. The detector response is built synthetically and plays the role of the Geant4-derived response matrix $R_{ij}$ of Sec.~\ref{subsec:detector_response}; because the reweighting acts on the recoil spectrum upstream of the response and treats $R_{ij}$ as a fixed input, the validation is indifferent to whether $R_{ij}$ is synthetic or obtained from a full detector simulation.

\subsection{Closure against an analytic reference}
\label{subsec:closure}

We construct a test problem in which the reconstructed-observable spectrum can be computed exactly, so that the reweighting prediction can be compared against a true reference rather than against a second Monte-Carlo estimate. We take a monoenergetic $^7$Be neutrino line ($E_\nu=0.862$~MeV) scattering on electrons. The Standard-Model recoil spectrum follows the closed-form cross section of Eq.~\eqref{eq:sm_cross_section}; an NSI modification shifts the chiral couplings $g_{L,R}\to g_{L,R}+\varepsilon_{ee}^{L,R}$ in that same expression, and a magnetic moment adds the $1/T$ term of Eq.~\eqref{eq:xs_mm}. For the detector response we adopt an analytic model with a light yield of $500$ photoelectrons per MeV, an ionization-quenching non-linearity that suppresses the low-energy light yield, and a Gaussian approximation to photoelectron-counting (Poisson) resolution, with variance equal to the mean, $\sigma_{N_p}=\sqrt{\langle N_p\rangle(T)}$, so that the conditional density $P(N_p\,|\,T)$ is known in closed form. These choices mimic the qualitative features of a realistic liquid-scintillator response (Sec.~\ref{subsec:detector_response}) -- a non-proportional light yield and finite resolution -- but are otherwise arbitrary: the synthetic response stands in for a Geant4-derived one, which makes no difference to the closure because the test probes the reweighting, not the response.

This construction provides a reference, computed independently of the reweighting machinery, that is effectively exact at the precision of interest. The reference (``direct'') prediction is obtained by folding the \emph{modified} recoil spectrum through $P(N_p\,|\,T)$ directly on a fine continuum grid; for the analytic inputs above this is the reconstructed-observable spectrum to numerical-quadrature precision, far below the deviations reported here. The ``reweighting'' prediction is produced by the method of Sec.~\ref{subsec:2d_reweight}: the SM recoil spectrum is binned, a single weight $r^{(k)}$ is applied per recoil bin, and the result is folded through the fixed response matrix $R_{ij}$ assembled from the same analytic $P(N_p\,|\,T)$. Because the response is identical for the two predictions and is independent of the new-physics parameters, the reweighting is mathematically exact in the continuum limit, and the dominant difference between the two predictions is the evaluation of the weight at a single representative energy per recoil bin. Comparing them therefore isolates exactly that bin-level approximation, with no Monte-Carlo fluctuation to obscure it.

\begin{figure}[!htb]
  \centering
  \includegraphics[width=\linewidth]{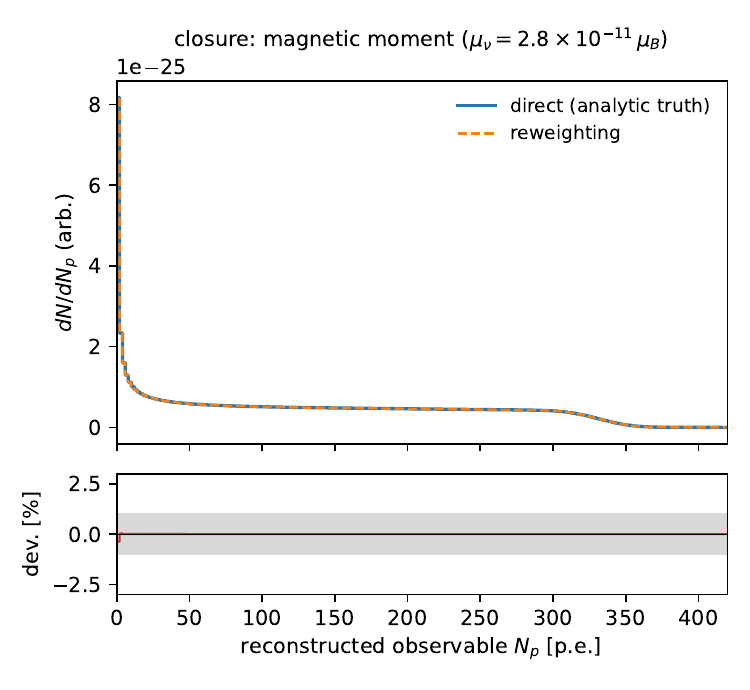}
  \caption{Closure test of the reweighting method against the analytic reference, for the magnetic-moment case ($\mu_\nu=2.8\times10^{-11}\,\mu_B$) -- the most stringent test, since the $1/T$ cross section varies most steeply within a recoil bin. \emph{Top:} reconstructed-observable ($N_p$) spectrum from the direct continuum fold (solid) and from the response-matrix reweighting (dashed) at the framework's $2000$-bin recoil grid; the two curves are indistinguishable. \emph{Bottom:} bin-by-bin relative difference, below $10^{-5}$ with no systematic trend.}
  \label{fig:closure}
\end{figure}

Figure~\ref{fig:closure} shows the comparison for the magnetic-moment case, the most stringent test because the $1/T$ cross section [Eq.~\eqref{eq:xs_mm}] varies most steeply within a recoil bin. At the framework's recoil binning ($2000$ uniform bins in $T$) the reweighting prediction reproduces the analytic reference with an integral deviation below $10^{-6}$ over the analysis region of interest ($50<N_p<400$ photoelectrons), and a bin-by-bin residual below $10^{-5}$ with no systematic trend. The NSI case ($\varepsilon_{ee}^{L}=\varepsilon_{ee}^{R}=0.1$) agrees equally well. This near-exact agreement is expected: for a parameter-independent response the reweighting is exact in the continuum, so the residual is purely a discretization effect.

\begin{figure}[!htb]
  \centering
  \includegraphics[width=\linewidth]{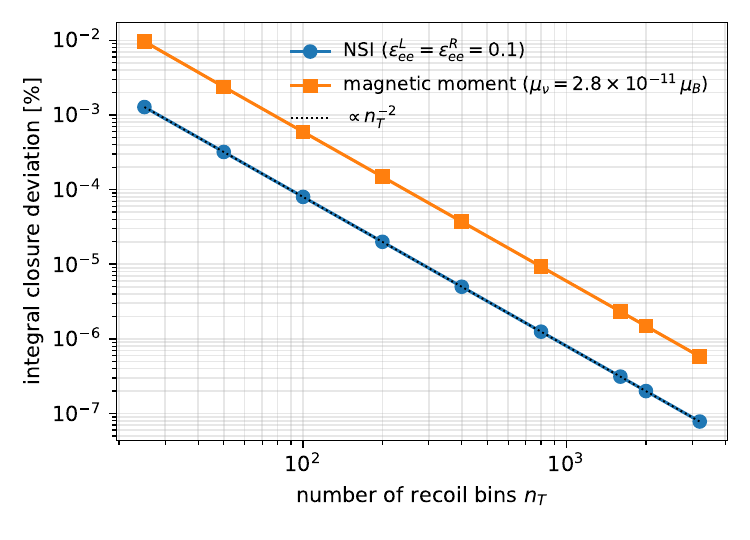}
  \caption{Integral closure deviation against the analytic reference as a function of the number of recoil bins $n_T$, for the NSI ($\varepsilon_{ee}^{L}=\varepsilon_{ee}^{R}=0.1$) and magnetic-moment ($\mu_\nu=2.8\times10^{-11}\,\mu_B$) cases. The deviation falls as $n_T^{-2}$ (dotted line), the expected midpoint-rule scaling, with no floor reached over the range tested. The framework's default $2000$-bin grid gives a deviation below $10^{-6}$.}
  \label{fig:closure_conv}
\end{figure}

The residual is controlled entirely by the recoil binning. Figure~\ref{fig:closure_conv} shows the integral closure deviation as a function of the number of recoil bins $n_T$: it falls as $n_T^{-2}$, the expected scaling of the single-point (midpoint) evaluation of the weight, with no floor reached over the range tested. Even a coarse $25$-bin grid closes to better than $0.01\%$ -- $0.0013\%$ for the NSI case and $0.0097\%$ for the steeper magnetic-moment weight. The magnetic-moment deviation is about seven times the NSI deviation at fixed binning, reflecting its steeper variation within a bin, but it remains negligible at any binning the framework uses. The $2000$-bin grid therefore holds the discretization error far below the precision of any foreseeable measurement, and the error budget is set by a single, freely adjustable parameter.

One caveat carries over to a realistic application, where the baseline is a finite Monte-Carlo sample rather than an analytic spectrum. Reweighting from a fixed sample is reliable only where that sample populates the support of the target spectrum. The magnetic-moment cross section scales as $1/T$ and is largest at low recoil energy, where a sample generated under the SM hypothesis is sparse; there the per-event weights grow and the effective sample size $N_{\rm eff}=(\sum_e w_e)^2/\sum_e w_e^2$ falls below the nominal count. A magnetic-moment-dominated analysis, whose sensitivity comes from the low-recoil bins, therefore requires the baseline sample to be sized -- or importance-sampled with a flatter recoil spectrum -- so that $N_{\rm eff}$ remains adequate there. This is a property of the baseline sample, not of the reweighting itself, and is absent in the analytic closure above; we note it because it governs how the method should be deployed on production Monte Carlo.

\subsection{End-to-end demonstration: injection and recovery}
\label{subsec:demo}

To exercise the complete pipeline -- parameter update, reweighting, response folding, and likelihood evaluation -- repeatedly over a scan, we perform an Asimov injection--recovery test on the same analytic model. For each physics example we build an Asimov data set from the analytic reference at a chosen injected parameter value, then scan the parameter using the reweighting method, profiling an overall rate normalization at each scan point and mapping the resulting profile likelihood to a confidence interval through the asymptotic $\Delta\chi^2$ relation~\cite{Cowan:2010js}. This is the workload for which the per-call performance of Sec.~\ref{sec:performance} is relevant: a one-dimensional scan is a few hundred such evaluations.

\begin{figure}[!htb]
  \centering
  \includegraphics[width=\linewidth]{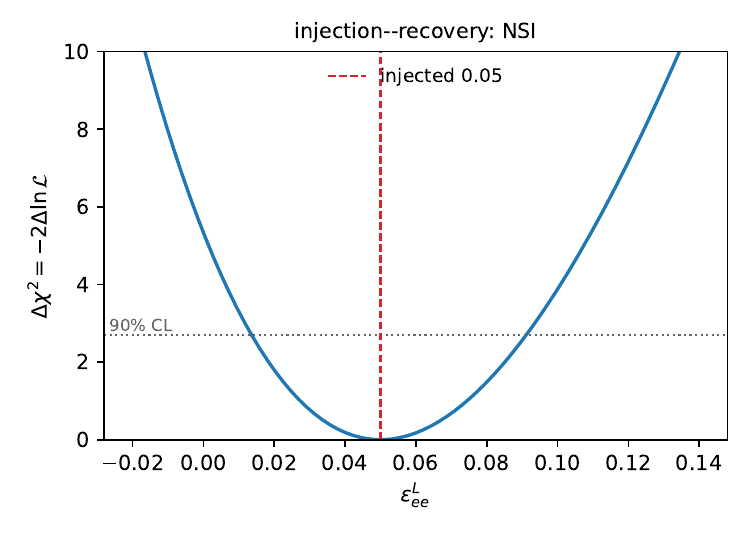}
  \caption{End-to-end injection--recovery for the NSI coupling $\varepsilon_{ee}^{L}$. Profile likelihood $\Delta\chi^2=-2\Delta\ln\mathcal{L}$ from a scan of synthetic Asimov data generated at the injected value $\varepsilon_{ee}^{L}=0.05$ (dashed line); the dotted line marks the $90\%$ confidence level (one degree of freedom). The minimum coincides with the injected value. Each scan comprises a few hundred likelihood evaluations of the kind timed in Table~\ref{tab:timing}.}
  \label{fig:demo}
\end{figure}

\begin{figure}[!htb]
  \centering
  \includegraphics[width=\linewidth]{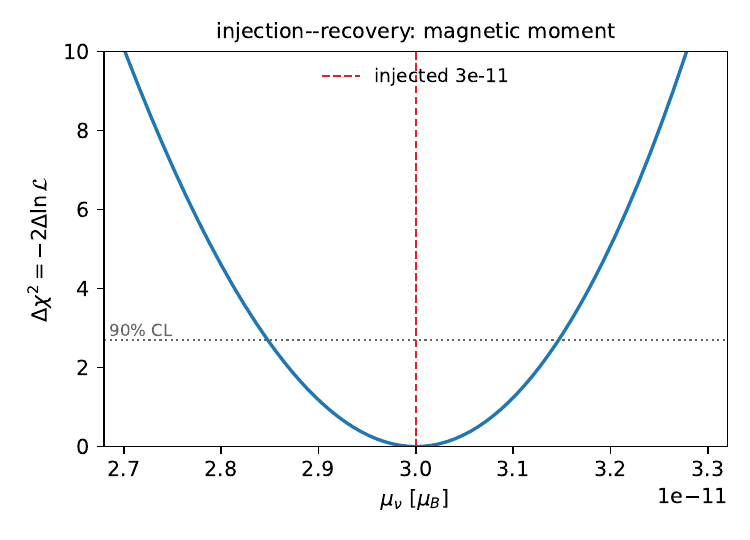}
  \caption{End-to-end injection--recovery for the neutrino magnetic moment $\mu_\nu$, from the same analytic model. Asimov data are generated at the injected value $\mu_\nu=3\times10^{-11}\,\mu_B$ (dashed line); the dotted line marks the $90\%$ confidence level. This scan exercises the steep $1/T$ reweighting whose bin-level accuracy is quantified in Sec.~\ref{subsec:closure}.}
  \label{fig:demo_mm}
\end{figure}

Figures~\ref{fig:demo} and~\ref{fig:demo_mm} show the resulting profiles. The reweighting method recovers the injected values -- $\varepsilon_{ee}^{L}=0.05$ and $\mu_\nu=3\times10^{-11}\,\mu_B$ -- with the injected point lying at the likelihood minimum in each case, confirming that the complete scan is internally consistent (the pipeline returns the injected value) for both the smooth NSI weight and the steep $1/T$ magnetic-moment weight. The width of each profile reflects only the assumed Asimov statistics of the synthetic data set and carries no physical meaning; these are demonstrations of the computational pipeline, not measurements. A physics analysis on real data, with its full treatment of systematic uncertainties, is beyond the scope of this methods paper.

\section{Conclusions and outlook}
\label{sec:conclusions}

We have presented a GPU-accelerated spectrum-reweighting framework for new-physics searches in solar neutrino--electron scattering. The method incorporates the parameter dependence of flavor-diagonal NSIs and an effective neutrino magnetic moment through bin-to-bin weights applied to recoil spectra derived from a baseline SM MC sample. By combining this reweighting procedure with a two-dimensional detector-response model derived from Geant4 simulation, the framework retains the detector response while avoiding the regeneration of detector-level samples at each point in parameter space.

The analysis chain is implemented with \textsc{Thrust} transformation--reduction primitives and organized into a portable software architecture that supports both CUDA and OpenMP back ends from a common source. In this implementation, the computation of recoil-energy spectra, detector-response propagation, and likelihood evaluation is expressed as a sequence of data-parallel transformations. This separation keeps the physics definitions independent of the hardware back end while allowing repeated likelihood evaluations to be accelerated on GPUs.

For the benchmarks tested here, one likelihood call takes ${\sim}52$~ms on a datacenter NVIDIA A30X GPU ($58\times$ over a single CPU thread, $2.5\times$ over the best 64-thread CPU) and ${\sim}87$~ms on a commodity RTX~3080~Ti, confirming interactive performance on hardware ranging from rentable consumer GPUs to datacenter accelerators. These timing results support the use of the framework for repeated scans and likelihood fits.

The scope of the framework is determined by the factorization assumption used throughout the paper. It applies to new-physics scenarios whose effects can be represented as bin-to-bin reweighting of a recoil spectrum while leaving the detector response for a recoil electron of energy $T$ unchanged. This includes the flavor-diagonal NSI and magnetic-moment cases considered here. Scenarios involving qualitatively different final-state particles, detector signatures, or flavor-changing structures beyond the cross-section treatment specified in this paper would require additional formalism and validation.

Several directions remain for future work. On the physics side, the framework can be extended to more general NSI flavor structures, additional new-interaction operators, and combined analyses of multiple detector technologies once the corresponding cross sections and response mappings are specified. On the computing side, the same design can be tested on newer GPU platforms and additional parallel back ends, with complete reporting of the software environment, compiler settings, timing protocol, and numerical reproducibility checks.
\begin{acknowledgments}
This work was supported by the National Key Research and Development Program of China
under Grant Nos.~2024YFE0110501 and 2024YFE0110504. Computing resources were provided
by the IHEP computing centre.
\end{acknowledgments}

\end{document}